\documentclass[%
 reprint,
 unsortedaddress,
 amsmath,amssymb,
 aps,prb,
]{revtex4-2}

\usepackage[T1]{fontenc}
\usepackage{times}
\usepackage{graphicx}
\usepackage{bm} 
\usepackage{bbm}
\usepackage{amsfonts, amsmath,amssymb, bm, graphicx}
\usepackage{siunitx}
\usepackage{ulem}
\usepackage{amssymb}
\usepackage[english]{babel}
\usepackage{lipsum}

\usepackage{dcolumn} 
\usepackage{bm} 
\usepackage[english]{babel}
\usepackage{amssymb,amsmath}
\usepackage{hyperref}
\hypersetup{
    colorlinks=true,
    linkcolor=blue,
    citecolor=blue,
    urlcolor=blue
}
\usepackage{color}

\usepackage{soul}
\def\vec#1{{\bm{#1}}}

\begin{document}
\preprint{AIP/123-QED}

\title{Nonreciprocal spin waves of helical magnetization states in CoFeB/NiFe bilayers}

\author{Claudia Negrete}
\affiliation{Departamento de F\'isica, Universidad Cat\'olica del Norte, Avenida Angamos 0610, Antofagasta, Chile}

\author{Omar J. Suarez}
\affiliation{Departamento de F\'{i}sica, Universidad de Sucre,  Sincelejo, Colombia}

\author{Attila K\'{a}kay}
\affiliation{Helmholtz-Zentrum Dresden~-~Rossendorf, Institute of Ion Beam Physics and Materials Research, Bautzner Landstr. 400, 01328 Dresden, Germany}

\author{Jorge A. Ot\'alora}
\email[Corresponding author: ]{jorge.otalora@ucn.cl}
\affiliation{Departamento de F\'isica, Universidad Cat\'olica del Norte, Avenida Angamos 0610, Antofagasta, Chile}

\date{\today}

\begin{abstract}
 We investigated the nonreciprocal spin-wave properties, including the frequency shift, of a helical equilibrium state in a versatile CoFeB/NiFe bilayer. Through an extension of the dynamic matrix formalism (developed in this work) to an arbitrary non-collinear configuration along a heterostructured multilayered system thickness, we explained the frequency shift via differences in the dynamic dipolar and interlayer exchange interactions arising from the distinct spin-wave mode profiles across the bilayer thickness for counterpropagating modes at the same wave vector. In contrast to recent literature wherein the frequency shift is attributed solely to the dipolar interaction, our results and explanations hereby presented involve a starring role of the interlayer exchange interaction not accounted in current literature. Furthermore, we also found a combination of large frequency shift values and sub-100 nm spin wave wavelengths that can be tuned or even enhanced with the twisting degree of the helical magnetization state by the application of the external field, and with the thickness of the NiFe sublayer, which might be highly relevant for magnonic applications. We validated our model and the physical mechanism that explains the frequency shift using recent simulations and experimental results.  
\end{abstract}

\maketitle

\section{Introduction}

The directional control of spin waves has become a central theme in modern magnonics due to its potential for low-energy and charge-free information transport and processing \cite{Kruglyak2010, Mahdi2013,Demokritov2018, KRUGLYAK2006191, LENK2011107}. In this context, non-reciprocity, characterized by an asymmetry in the dispersion relation such that  $f[k] \neq f[-k]$ \cite{Nembach213256,PhysRevLett.117.227203,PhysRevB.98.014403}, plays a fundamental role, as it enables unidirectional propagation and the design of functional devices such as magnonic diodes \cite{PhysRevX.5.041049, PhysRevApplied.14.034063}, insulators, and reconfigurable logic elements \cite{Adam2002, chen2006oriented}. This asymmetry can arise from various physical mechanisms, including dynamic dipolar interactions in Damon-Eshbach geometries  \cite{PhysRev.178.839, PhysRevB.66.132402,Flebusarticle,PhysRevB.104.184429}, structural symmetry breaking in thin films, and the Dzyaloshinskii Moriya interaction (DMI) induced by strong spin-orbit coupling in ferromagnet/heavy metal heterostructures \cite{PhysRevLett.118.147201,PhysRevB.106.014405, 10.21468/SciPostPhys.7.3.035}. Such mechanisms have been extensively explored both experimentally and theoretically in magnetic bilayers and multilayers, where manipulation of thickness and magnetic properties allows modulation of the magnitude of nonreciprocity.

In recent years, increasing attention has been paid to heterostructured multilayered planar systems with spatial variations in magnetic properties along the thickness, including synthetic antiferromagnets and multilayers with graded magnetization profiles \cite{PhysRevB.104.174417,Gallardo_2019,PhysRevB.104.174417,PhysRevB.31.4465,PhysRevB.102.184424}. In these systems, isotropic exchange coupling and collective dipolar interactions can lead to spin wave modes hybridization, surface localization, and modification of the dynamic character of the excitations \cite{Gallardo_2019,Fallarino_2018,Fallarino_2017}, leading to thickness-related caustic effects, the emergence of hybrid modes with locally inverted precession (anti-Lamor precession), and substantiated non-reciprocal properties. \cite{Heins2024,Gallardo_2019,PhysRevB.104.174417}

Recently, highly tunable nonreciprocal propagation in SmCo/Fe bilayers, known as spring-exchange systems, has been experimentally demonstrated and simulated using micromagnetic methods. In this bilayer, the in-plane easy axis anisotropy of the hard magnetic layer (SmCo) and the response of the soft magnetic layer (Fe) as in-plane deviations of its magnetization from the hard layer easy axis, generate a non-collinear magnetization profile along the thickness (so named here as the helical magnetic state) that allows a tunability of the nonreciprocity with the layer thicknesses and the applied magnetic field \cite{l93m-gb54,Taaev2024,PhysRevB.93.144409,Liu2008}. Within this framework, exchange spring magnets constitute a particularly attractive platform, as they combine the power of two seemingly separate worlds: spring exchange valves and a multilayered non-reciprocal platform for magnonic applications. These nanostructures, typically composed of a hard magnetic phase and a soft phase coupled by exchange, exhibit a helical or gradually rotated equilibrium state along the thickness due to the competition among anisotropy, exchange, and Zeeman interactions\cite{Fullerton1999,Magnus2016}. Originally proposed for high-density magnetic storage applications, these systems also offer a natural environment for investigating spin wave propagation in media with spatially distributed non-collinear magnetization \cite{Haldar2014BrillouinLS,PhysRevB.63.174415}. Analyzing non-reciprocity in such structures allows us not only to clarify the origin of the dispersion relation asymmetry but also to evaluate its potential for thickness-controlled modal engineering in next-generation magnonic devices. To date, no theoretical framework with analytical support has been found that systematically explains how helical magnetization, magnetization variation, and the applied field determine non-reciprocity in spin waves.

\begin{figure}[h]
\includegraphics[scale=0.14]{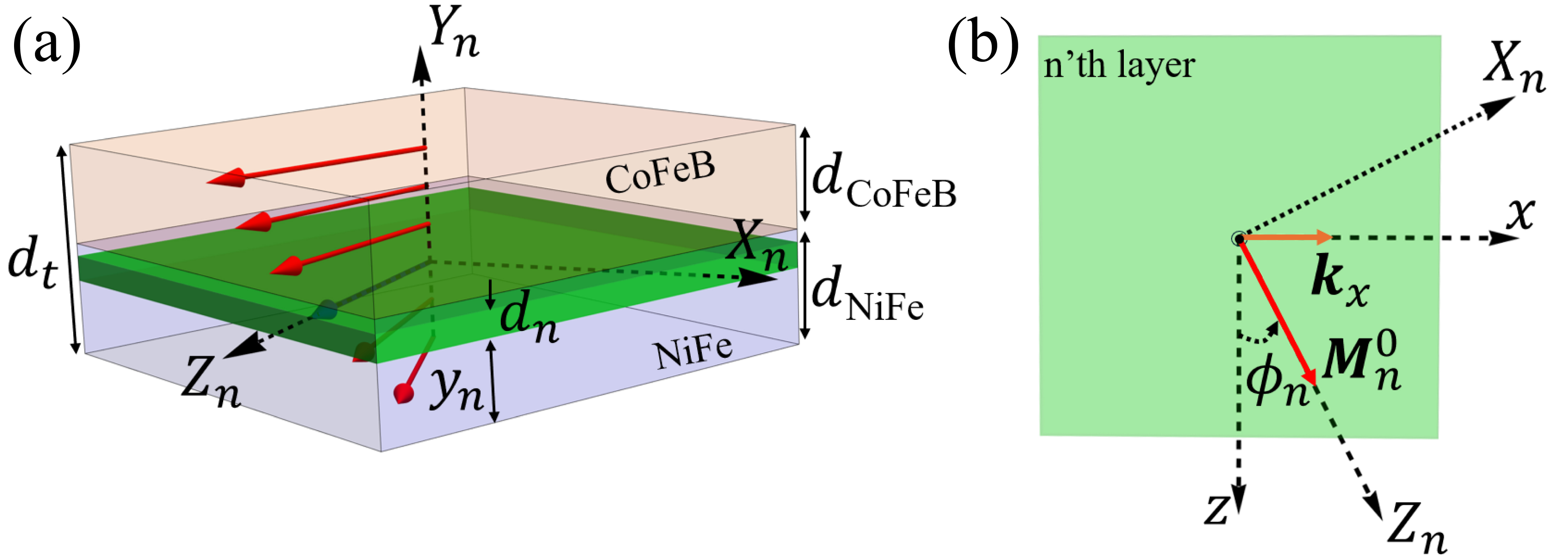}
\caption{Illustration of CoFeB/NiFe bilayer in a helical magnetization state with a total thickness $d_t$, NiFe thickness $d_{\text{Py}}$, and CoFeB thickness $d_{\text{CoFeB}}$. (a) Illustration of the system subdivided into N sublayers, where the n'th layer is positioned at the $y_n$, with thickness $d_n$, saturation magnetization $M_s^n$, and equilibrium magnetization $\boldsymbol{M}_n^0=M_{s_n}\hat{Z}_n$, where $(X_n,Y_n,Z_n)$ are local coordinates. (b) Illustration of the n'th layer with its equilibrium angle $\phi_n$ regarding the global coordinate $z$. Global coordinates are denoted by $(x,y,z)$ and the spin-wave wave vector is $\boldsymbol{k}_x=k_x\hat x$.}
\label{fig:system}
\end{figure}

In this work, we extend the dynamic matrix formalism to describe spin-wave propagation in a multilayer with an in-plane equilibrium magnetization that twists layer by layer along the multilayer thickness. We apply our model to understand the non-reciprocal spin wave transport properties of a CoFeB/NiFe bilayer exchange spring with a helical equilibrium magnetization profile along its thickness, as seen in FIG. \ref{fig:system}. We choose this system due to its versatility for exchange spring applications \ cite {Tacchi2014,Stenning2015,Gubbiotti2021} and magnonic diodes \cite{Heins2024,GrassiDiode2020}. Starting from a multilayer model, we determine the equilibrium configuration and subsequently calculate the spin-wave dispersion relation and its directional asymmetry as a function of bilayer thickness and applied field. Special attention is paid to the layer-resolved dynamic mode profiles, where exchange and dipolar interactions play a significant role in explaining the frequency shift of the nonreciprocal dispersion relation. Our analytical formalism was validated with micromagnetic simulation and experimental results previously published by other authors on the SmCo/Fe exchange spring bilayer \cite{l93m-gb54}, which can be found in the supplementary material Section II\cite{SupMat}, as well as details on the mathematical derivations of the analytical equations presented in here and shown in supplementary material Section I\cite{SupMat}.

\smallskip{}

\section{Model}
\label{model}

The system under consideration is modeled under the micromagnetism continuum theory and the dynamic matrix method (DMM). According to the later, the CoFeB/NiFe bilayer is assumed to be a multilayer stack of N-sublayers, where the $\mathrm{n'th}$ sublayer (one of the layers of the total number of layers) is magnetically described by a magnetization vector $\vec{M}_\mathrm{n}$, an exchange length $l_{\mathrm{ex}_n}$, a stiffness constant $A_{n}$ and a saturation magnetization $M_{s_{n}}$, as shown in FIG \ref{fig:system}. These assumptions, along with the following considerations on the nth layer, allow for achieving the in-plane helical magnetization states, as illustrated in FIG. \ref{fig:system}(a). We assume the following: (i) an in-plane and homogeneous applied magnetic field $\mathbf{H}=H \hat z$, (ii) an uniaxial anisotropy interaction featured by a constant $K_{n}^{u}$ that induce a magnetization tendency to line-up along the $\hat z$ axis, (iii) an interlayer exchange interaction featured by the integral exchange constant $J_{n,p}$ between the $\mathrm{n'th}$ and $\mathrm{p'th}$ sublayers, and (iv) an intralayer exchange interaction featured by the exchange length $l_{\mathrm{ex}_n}$. Under this scenario, the helical magnetization state can be stabilized and featured by the equilibrium magnetization $\mathbf{M}_n^0$ and the magnetization angle $\phi_n$, as illustrated in FIG. \ref{fig:system}(b). These angles are obtained by solving Brown's equation, as we will see later. The spin-waves properties as the dispersion relation $\omega(\mathbf{k})$ of the helical equilibrium state (also referred here as the magnetization dynamics) will be studied for the specific case of a wave vector $\mathbf{k}=\mathbf{k}_x=k_x \hat x$ pointing along the global coordinate $x$, as shown in  FIG. \ref{fig:system}(b). The angular spin-wave frequency $\omega$ as a function of the wave vector $\mathbf{k}_x$ will be addressed next by linearizing the magnetization dynamical equation around the helical equilibrium state. 

According to the micromagnetic continuum theory, the equilibrium states and the magnetization dynamics of the bilayer can be modeled by the Landau-Lifshitz (LL) equation. Since the system is divided into N layers, we end up with N coupled LL equations, one for each layer. In this sense, the LL equation of the $\mathrm{n'th}$ layer is given as

\begin{equation}
    \dot{\mathbf M}_n[\mathbf r,t]=-\mu_0\gamma\mathbf{M}_n[\mathbf r, t]\times\mathbf{H}_n^e[\mathbf r,t]
    \label{eq:LLG}
\end{equation}

\noindent where  $\gamma$ is the magnitude of the gyromagnetic ratio, $\mathbf{H}_{n}^{e}[\mathbf{r},\,t]$  is the effective magnetic field acting on the $\mathrm{n'th}$ layer, and $\mathbf{M}_{n}[\mathbf{r},\,t]$ is the magnetization field. 
The spin-wave dispersion relation and the equilibrium helical state can be studied by linearizing each of the $\mathrm{n'th}$ LL-equation around the equilibrium magnetization $\mathbf{M}_n^0$. To do so, the magnetization field and the effective field are linearly expanded as $\mathbf{M}_{n}[\mathbf{r},\,t]=\mathbf{M}_{n}^{0}+\mathbf{m}_{n}[\mathbf{r},\,t]$ and $\mathbf{H}_{n}^{e}[\mathbf{r},\,t]= \mathbf{H}_{n}^{0e}+\mathbf{h}_{n}^{e}[\mathbf{r},\,t]$, respectively, where $\mathbf{M}_{n}^{0}$ and $\mathbf{H}_{n}^{0e}$ are the magnetization and the effective field at equilibrium, respectively, and the terms $\mathbf{m}_{n}[\mathbf{r},\,t]$ and $\mathbf{h}_{n}^{e}[\mathbf{r},\,t]$ are dynamic perturbations. We assume that the dynamic magnetic perturbation and dynamic magnetic field perturbation have the form $\mathbf{m}_{n}[\mathbf{r},t] =  \int d^3\mathbf{k}\,e^{i(\omega t-\mathbf{k}\cdot \mathbf{r})} \mathbf{m}_n[\mathbf{k}]$, and $\mathbf{h}_{n}[\boldsymbol{r},t] =  \int d^3\mathbf{k}\,e^{i(\omega t-\mathbf{k}\cdot \mathbf{r})} \mathbf{h}_n[\mathbf{k}]$ respectively, where $\mathbf{m}_n[\mathbf{k}]$ and $\mathbf{h}_n[\mathbf{k}]$ are the dynamic magnetization and dynamic magnetic field amplitude vector, respectively, written in the spin-wave wave vector $\mathbf{k}$ space, and $\omega$ is the angular resonance frequency of the spin-waves. After linearizing Eq. \eqref{eq:LLG}, we arrive to the following two equations: 

\begin{eqnarray}
\mathbf{M}_{n}^{0}\times\mathbf{H}_{n}^{0e} & = & 0\label{eq:EQUATION BROWN},\\
\omega\,\mathbbm{m}[\mathbf{k}] & = & \mathbb{N} \mathbbm{m}[\mathbf{k}]\label{eq:LINEALIZA LL},
\end{eqnarray}

\noindent where equation EQ. \eqref{eq:EQUATION BROWN} is the Brown's equation, whose solutions return the equilibrium state of the system (in our case, a solution for the equilibrium angles $\phi_1,...,\phi_{N}$ that define the helical states); whereas the equation EQ. \eqref{eq:LINEALIZA LL} represents the eigenvalue equation, being $\mathbb{N}$ the so-called dynamic matrix, whose dimensions are $2N\,\times\,2N$, with $N$ the number of sublayers and  $\mathbbm{m}[\mathbf{k}]$ the eigenvector with a transpose given as $\mathbbm{m}[\mathbf{k}]^{T}=\left\{ \tilde m_{\text{X}_{1}}[\mathbf{k}],\text{...},\tilde m_{\text{X}_{N}}[\mathbf{k}],\tilde m_{\text{Y}_{1}}[\mathbf{k}],\text{...},\tilde m_{\text{Y}_{N}}[\mathbf{k}]\right\}$.

The Brown's equation for the $\mathrm{n'th}$ layer Eq.(\ref{eq:EQUATION BROWN}) in its explicit form is given as

\begin{align}
\sin\left(\phi_H-\phi_n \right) &+ \frac{H_{n,n+1}^{\text{inter}}}{H} \sin\left( \phi_n - \phi_{n+1} \right) \notag \\
&+ \frac{H_{n,n-1}^{\text{inter}}}{H} \sin\left( \phi_n - \phi_{n-1} \right) \notag \\
&- \frac{\tilde{H}_{n}^u}{H} \cos\left( \phi_n \right) \sin\left( \phi_n \right) = 0.
\label{Eq:BrownEqExplicit}
\end{align}

\noindent The first term corresponds to the Zeeman interaction where a magnetic field $H$ is applied in-plane at an angle $\phi_H$ measured from the $z$ axis. The second and third terms contain the interlayer exchange that accounts for the coupling between the $\mathrm{n'th}$ layer and its two neighbouring layers $n\pm1$. Such coupling are strengthen by the interlayer exchange field $H_{n,p}^{\text{inter}} = -J_{n,p}/(2 M_{s_n} \mu_0 d)$. The last term corresponds to the uniaxial anisotropy of the n'th layer, where $\tilde{H}_{n}^{\text{u}}=2| K_n^{\text{u}}|/(\mu_0 M_{s_n})$ is the strength of the anisotropy field. The dipolar and intralayer exchange interactions do not appear in the Brown equation. 

The solution for the set of angles $\phi_1,...,\phi_{N}$ is obtained by solving the system of N equations resulting from applying EQ. \eqref{Eq:BrownEqExplicit} at $n=1,...,N$.   

The dynamic matrix contains all the information about the static and dynamic effective fields acting on the magnetization field. It is recorded as the additive contribution of the dynamic tensors associated to the dipolar, exchange, uniaxial anisotropy, and Zeeman interactions:  $\mathbb{N}=\mathbb{N}^{\mathrm{dip}}+\mathbb{N}^{\mathrm{inter}}+\mathbb{N}^{\mathrm{intra}}+\mathbb{N}^{u}+\mathbb{N}^{Z}$, where the dynamic tensor contribution $\mathbb{N}^{\sigma}$ is given as

\[
\mathbb{N}^\sigma =
\begin{pmatrix}
\mathbb{N}^{XX,\sigma} & \mathbb{N}^{XY,\sigma} \\
\mathbb{N}^{YX,\sigma} & \mathbb{N}^{YY,\sigma}
\end{pmatrix}
\]

\noindent with $\sigma\in\left\{\mathrm{dip,  inter, intra, u, Ze}\right\}$. The dynamic matrix blocks $\mathbbm{N}^{q,\sigma}$  with $q\in\{XX,XY,YX,YY\}$ are $N\times N$ matrices where the entries $N_{np}^{q,\sigma}=[\mathbbm{N}^{q,\sigma}]_{np}$ are explicitly defined in the supplementary material  Section IB.\cite{SupMat}

 The dispersion relation $\omega[\mathbf{k}]$ and the eigenvectors $\mathbbm{m}[\mathbf{k}]$ are obtained by solving the eigenvalue EQ. \eqref{eq:LINEALIZA LL}.

\section{Results and Discussion}

In the following, we assume a planar CoFeB/NiFe bilayer that is subdivided into N sublayers. For simplicity and convergence of our calculations, each nth layer has the same thickness, $d=1$ nm. The CoFeB layer saturation magnetization,  stiffness constant, thickness, and uniaxial anisotropy constant are set to $M_s^{\mathrm{CoFeB}}$ = \SI{1270}{\kilo\ampere/\meter} $A_{\mathrm{CoFeB}}$ = \SI{17}{\pico\joule/\meter}, $d_{\mathrm{CoFeB}}=$\SI{25}{\nano\meter}, and $K_{\mathrm{CoFeB}}=500$ kJ/m$^3$, respectively. The NiFe layer saturation magnetization, stiffness constant, and uniaxial anisotropy constant are set to $M_s^{\text{NiFe}}$ = 845 $\text{kA/m}$, $A_{\text{NiFe}}=12.8$ $\text{pJ/m}$, and $K_{\text{NiFe}}=0$, respectively. The NiFe thickness and the applied magnetic field magnitude are free parameters that we run over a set of achievable experimental values to stabilize the equilibrium helical state and thereby study its magnetization dynamics. The applied magnetic field orientation is set to $\phi_H=0$, with $H\geq 0$.
In the following sections, we proceed to describe the helical equilibrium states by solving EQ. \eqref{Eq:BrownEqExplicit} as the thickness of the NiFe changes and for different magnetic field values $H$. Thus, we proceed with describing the dispersion relation and spin wave modes by solving EQ. \eqref{eq:LINEALIZA LL} as the helical equilibrium states are stabilized.

\subsection{Equilibrium states}

\begin{figure}[h]
\includegraphics[scale=0.19]{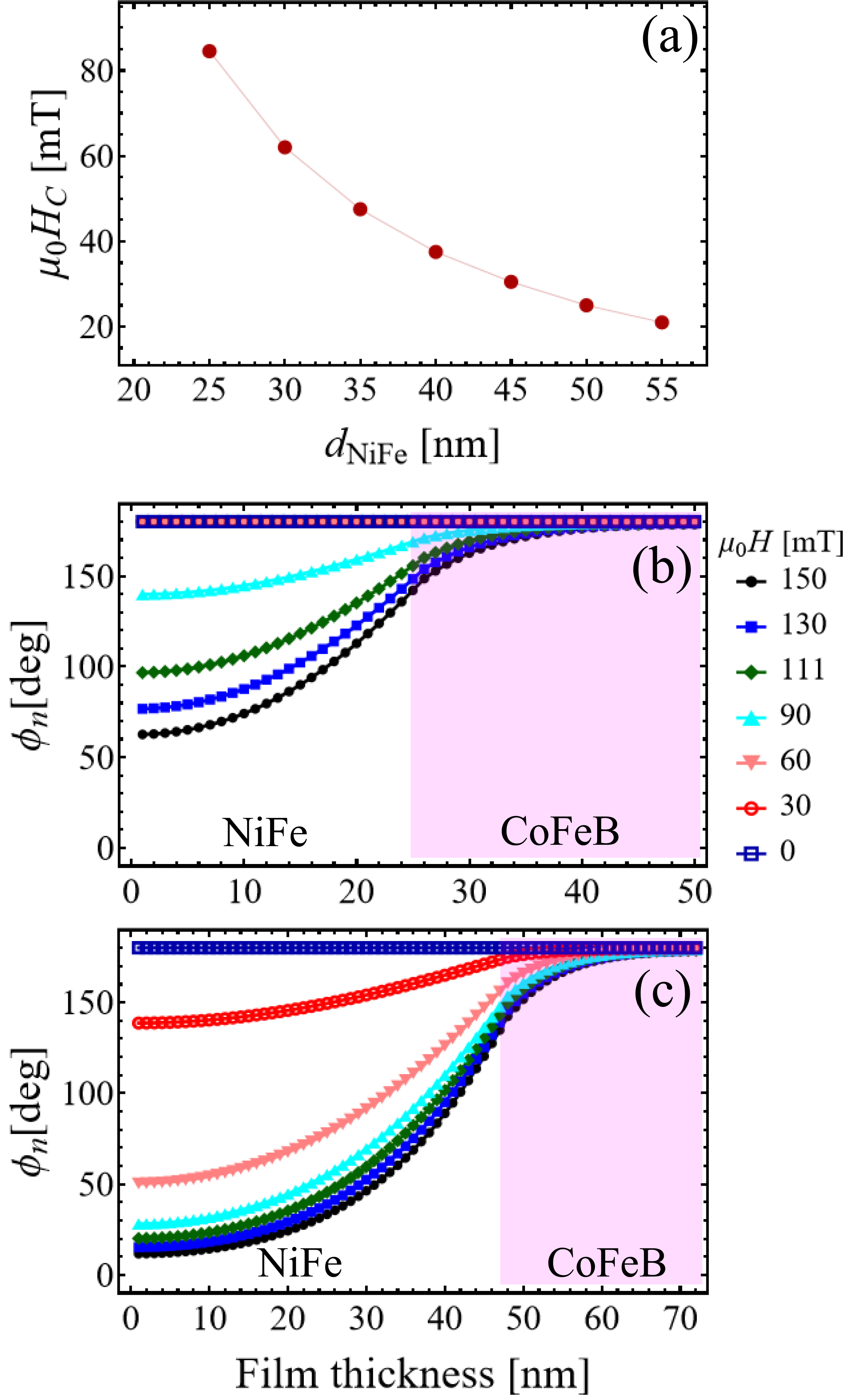}
\caption{(a) Coercive field as function of Permalloy thickness $d_{\text{NiFe}}$. Helical equilibrium states of the bilayer CoFeB/NiFe, with CoFeB (NiFe) thickness $d_{\text{CoFeB}} = 25 \, \text{nm}$ ($d_{\text{NiFe}} = 25 \, \text{nm}$), and $K_{\text{CoFeB}} = 500 \, \text{kJ/m}^3 $. The equilibrium helical states are shown in (b) for $d_{\text{NiFe}} = 25 \, \text{nm}$ and (c) for $d_{\text{NiFe}} = 47 \, \text{nm}$.}
\label{Helical_States_Hcoercive}
\end{figure}

FIG. \ref{Helical_States_Hcoercive}(a) presents the coercive field $H_C$ of the CoFeB/NiFe bilayer as a function of the thickness of the NiFe layer. The equilibrium magnetization states are also shown for two representative NiFe thicknesses, $d_{\text{NiFe}} = 25 \, \text{nm}$ and $d_{\text{NiFe}} = 47 \, \text{nm}$. In these results, the CoFeB thickness was kept constant at $d_{\text{CoFeB}} = 25 \, \text{nm}$. 

At zero applied magnetic field $H=0$, the equilibrium magnetization state is homogeneous, and for our convenience is assumed pointing along $-\hat z$ axis, i.e., $M_n^0=-M_{s_n}\hat z$. For an applied magnetic  $H>0$, the homogeneous state remains as long as the condition $H<H_C$ is satisfied, where $H_C$ is the coercive field. For a larger field magnitude $H>H_C$, the helical magnetization is stabilized. In our case, the coercive field $H_C$ defines the field onset where the NiFe layer magnetization loses the uniform configuration and turns into the helical state, as illustrated in FIG \ref{fig:system}(a). 

FIGs \ref{Helical_States_Hcoercive}(b) and (c) show in detail the helical states through the angle $\phi_n$ as a function of the layer position $y_n$ in the thickness of the film. The helical equilibrium states presented herein are in agreement with literature \cite{l93m-gb54}; moreover, they are expected in multilayered magnetic materials, wherein neighboring layers are coupled among them by the interlayer exchange interaction of the type taken here (see Supplementary Material Section IA3\cite{SupMat}). 

\subsection{Dispersion relation}

In the previous sections, we showed our results and discussion on the helical equilibrium states of the CoFeB/NiFe bilayer.  The next step is dedicated to the dispersion relation calculated with the dynamic matrix method EQ. \eqref{eq:LINEALIZA LL} with all relevant interactions explained in the Model section \ref{model}. In what follows, we calculate and analyze the dispersion relation of a CoFeB/NiFe bilayer in a helical magnetization state.  
Our results are obtained for different applied magnetic field magnitudes and different NiFe thicknesses. We will analyze the dispersion relations of the fundamental and first-order modes, focusing on the non-reciprocal frequency shifts and spin-wave orbital profiles along the thickness to explain the obtained results.

\begin{figure*}[t]
\centering
\includegraphics[scale=0.105]
{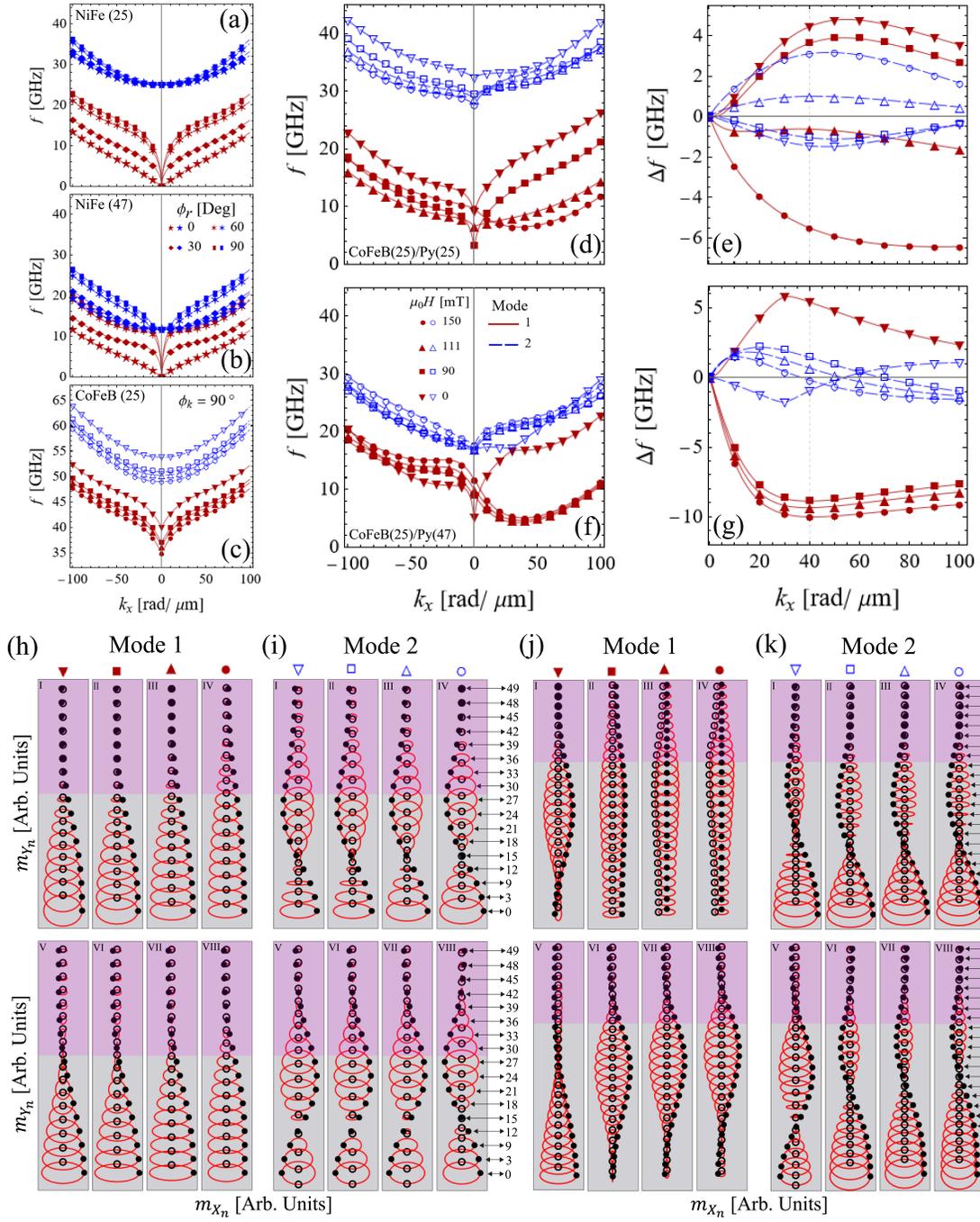}
\caption{Dispersion relation of an insulated and homogeneously saturated NiFe layer at different propagation directions $\phi_r$ and zero applied magnetic field $\mu_0 H=0$ mT of thickness (a) $d_{\text{NiFe}}=25$ nm and (b) $d_{\text{NiFe}}=47$ nm, respectively.  (c) Damon-Eschbach dispersion relation of an insulated and homogeneously saturated CoFeB layer of thickness $d_{\text{CoFeB}}=25$ nm and in-plane uniaxial anisotropy $K_{\text{CoFeB}}$ = 500  $\text{kJ/m}^3$, at different magnetic fields applied in $-\hat z$ direction. (d) Dispersion relation and (e) frequency shift $\Delta f=(f[k_x]-f[-k_x])$ of CoFeB/NiFe system with  $d_{\text{NiFe}} = 25 \, \text{nm}$. (f) Dispersion relation and (g) frequency shift $\Delta f=(f[k_x]-f[-k_x])$ of CoFeB/NiFe system with  $d_{\text{NiFe}} = 47 \, \text{nm}$. (h-i) and (j-k) Mode profiles at $|k_x| =40$ rad$/\mu$m of and a set of applied field $\mu_0 H\in\{0, 90, 111, 150\}$ mT for $d_{\text{NiFe}} = 25$nm and $d_{\text{NiFe}} = 47$ nm, respectively, where top (bottom) line of modes are for $k_x=40$ rads$/\mu$m ($k_x=-40$ rads$/\mu$m).}
\label{Fig_Dispersion_relation}
\end{figure*}

Our following discussion is divided into two main parts. In the first part, we discuss general aspects of the CoFeB/NiFe bilayer dispersion relation (see FIG. \ref{Fig_Dispersion_relation}(d,f)), including the origin of the dispersion's frequency range and the modes' hybridization at larger thicknesses. It will be done by comparing the dispersion relation between the CoFeB/NiFe bilayer and the individual CoFeB and NiFe layers.  In the second part, the frequency shift of the non-reciprocal dispersion relation is discussed and explained by analyzing the dipolar and interlayer exchange interactions involved in the spin-wave modes profiles along the bilayer thickness.

\begin{figure*}[ht!]
\includegraphics[scale=0.19]{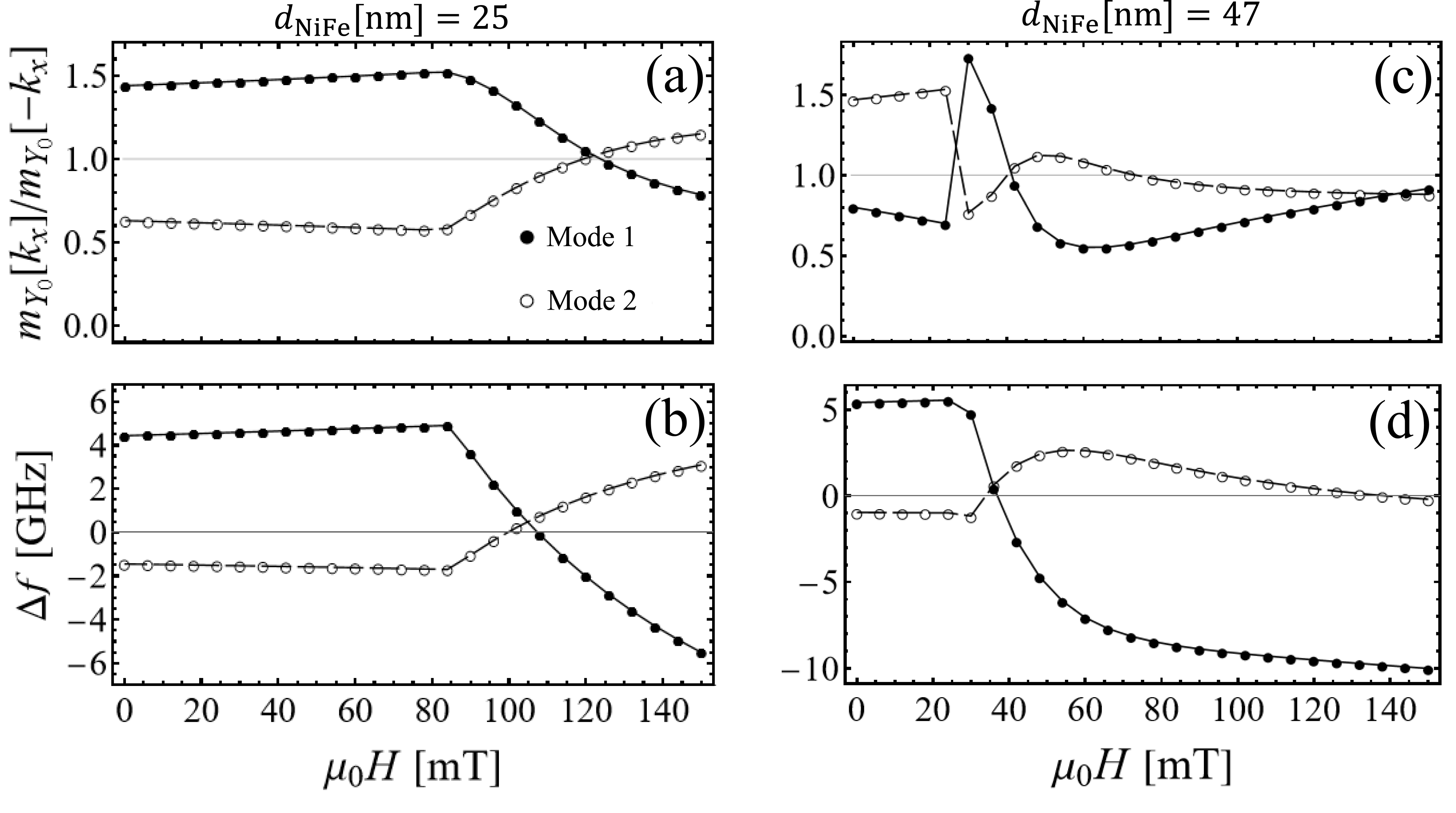}
\caption{(a,c) Surface charge ratio of the CoFeB/NiFe bilayer bottom surface and (b,d) its frequency shift as a function of the applied magnetic field. (a) and (b) corresponds to a NiFe sublayer thickness $d_{\text{NiFe}}=25$ nm, and (c) and (d) for $d_{\text{NiFe}}=25$ nm. A wave vector magnitude $|k_x|=40$ rads$/\mu$m was taken for calculations.}
\label{DensimyvsH}
\end{figure*}

The frequency range of the CoFeB/NiFe dispersion relation can be understood by comparing it with the frequency ranges of the single CoFeB and NiFe layers. In this sense, we consider the CoFeB and NiFe layers and assign to them a homogeneous magnetization state based on rough approximations of the helical equilibrium state of the CoFeB/NiFe bilayer. The structure of the helical equilibrium state in the CoFeB/NiFe bilayer is such that the magnetization of the NiFe layer rotates along its thickness, whereas the magnetization of the CoFeB layer is mostly antiparallel to the applied field, except for a rotation nearby the interface with the NiFe layer. Therefore, we base in the almost homogeneous magnetization of the CoFeB in the bilayer to assign an in-plane homogeneous state to the single CoFeB layer, with a constant relative orientation with respect to $\mathrm{k}_x$ given by $\phi_r=90$ Deg, which gives a dispersion relation as shown in FIG. \ref{Fig_Dispersion_relation}(c). Assigning the magnetization state to the single NiFe layer is a little subtle. The degree of twisting of the magnetization rotation of the NiFe layer in the bilayer depends on the applied magnetic field strength (see FIGs. \ref{Fig_Dispersion_relation}(a,b)), therefore, the relative orientation between the wave vector $k_x$ and the magnetization varies along the NiFe layer thickness in the bilayer. To account for this relative orientation in the dispersion relation of a single NiFe layer with a homogeneous magnetization, we calculate its dispersion relation at different wave vector orientations defined by the relative angle $\phi_r$ and at zero applied field $H=0$. Despite being a rough approximation, it shows that the dispersion relation frequency range of the CoFeB/NiFe bilayer mostly (poorly) matches that of the single NiFe (CoFeB) layer. Consequently, one can expect that the dynamic magnetization oscillates with a much larger amplitude in the NiFe layer than in the CoFeB layer in the bilayer, as observed and shown in FIGs. \ref{Fig_Dispersion_relation}(h-k). 

Comparing dispersion relations between the CoFeB/NiFe bilayer and the CoFeB and NiFe single layers is also helpful for understanding some general features of hybridization in the bilayer's spin wave modes. We present two cases, the CoFeB(25)/NiFe(25) bilayer without mode hybridization in the dispersion relation, as shown in FIG. \ref{Fig_Dispersion_relation}(d) and the CoFeB(25)/NiFe(47) bilayer with modes hybridization as shown in \ref{Fig_Dispersion_relation}(d). In the former case, the lack of mode hybridization is consistent with the lack of hybridization in the dispersion relation of the single NiFe layer \ref{Fig_Dispersion_relation}(a), wherein the CoFeB plays a side role due to its larger dispersion relation frequency range. In the latter case, the modes' hybridization in the CoFeB/NiFe bilayer dispersion relation FIG. \ref{Fig_Dispersion_relation}(f) occurs only at cero applied field $H=0$ and $k_x\approx 30 \text{ rads/}\mu\text{m}$, where the bilayer has an homogeneous equilibrium magnetization state. This is consistent with the modes' hybridization of the single NiFe layer shown in Fig \ref{Fig_Dispersion_relation}(b) at $\phi_r=90$ Deg and $k_x\approx 10 \text{ rads/}\mu\text{m}$. One could also expect modes hybridization at $k_x<0$ in the bilayer dispersion relation in consequence to the modes hybridization at $k_x\approx -10\text{ rads/}\mu\text{m}$ of the single NiFe layer. However, the lack of hybridization at $k_x<0$ in the bilayer is due to the fact that the CoFeB frustrates the heterosymmetrical condition for counter-propagating modes, i.e., the CoFeB tends to frustrate dynamic oscillations in its body and nearby the interface with the NiFe, avoiding establishing conditions for heterosymmetry, therefore, breaking magnetic interaction symmetries that lead to hybridization at $k_x<0$ as shown in FIG. \ref{Fig_Dispersion_relation}(f).

The non-reciprocal dispersion relation of the CoFeB/NiFe bilayer is quantified here by the frequency shift $\Delta f[k_x]=f[k_x]-f[-k_x]$, as shown in FIGs. \ref{Fig_Dispersion_relation}(e,g). One can see the behavior of $\Delta f$ v.s $k_x$ for the fundamental mode (mode 1) and first order mode (mode 2) at four different applied fields $H$ and two NiFe thicknesses. The magnetic field values were chosen such that, one can see the evolution of the frequency shift as the equilibrium magnetization turns from the homogeneous state to a helical state with an accentuated twist of the magnetization field in the NiFe layer at larger magnetic field values (see FIGs. \ref{Helical_States_Hcoercive}(b,c) at $\mu_0H\in\{0,90,110,150\}$ mT). The frequency shift as a function of the wave vector $k_x$ and the applied magnetic field $H$ can be understood by analyzing the effects on magnetic interactions arising from the asymmetry in the orbital distribution along the thickness of counterpropagating spin waves at the same wave vector.\cite{negrete2026exchange} FIG. \ref{Fig_Dispersion_relation}(h-k) shows the spin wave orbit distribution for modes 1 and 2 at four different applied magnetic fields and a representative wave vector magnitude $|k_x|=40$ rads$/\mu$m, being FIG. \ref{Fig_Dispersion_relation}(h,i) the spin wave orbit distribution for mode 1 and FIG. \ref{Fig_Dispersion_relation}(j,k) for mode 2. Notice that the top(bottom) panel in \ref{Fig_Dispersion_relation}(h-k) shows the spin wave modes orbits for $k_x=+40$ rads$/\mu$m ($k_x=-40$ rads$/\mu$m). 

In terms of interactions, the frequency shift is dominated by the interlayer exchange interaction when the counterpropagating spin wave modes' profiles at the same wave vector magnitude differ. Under these conditions, the interlayer exchange interaction contributes to the frequency shift by over two to three orders of magnitude more than any other interaction, including the dipolar interaction (in this work, we are not considering the Dzyaloshinskii?Moriya interaction). Exceptionally, the dipolar interaction can compete or even be dominant contributing interaction under the condition where the counterpropagating spin wave modes' profiles are very similar or identical to each other \cite{negrete2026exchange}. 
In the following, we use these arguments to analyze the frequency shift observed in two CoFeB/NiFe bilayers: the first with 25 nm NiFe thickness and the second with 47 nm NiFe thickness. We base our analysis on the spin wave orbit distribution of counterpropagating spin waves at the representative case of $|k_x|=40$ rads$/\mu$m shown in FIG. \ref{Fig_Dispersion_relation}(h-k), and analyzing the interlayer exchange and dipolar interactions to explain the frequency shift as a function of the applied magnetic field. As explained next, we see that the frequency shift of the CoFeB(25)/NiFe(25) of modes 1 and 2 is mostly dominated by the dipolar interaction, whereas the interlayer exchange interaction plays a fundamental role in both modes in the CoFeB(25)/NiFe(47). 

Let's start with analyzing the frequency shift of the CoFeB(25)/NiFe(25) bilayer. As seen in the upper ($k_x=+40$ rads$/\mu$m) and lower ($k_x=-40$ rads$/\mu$m) panels in FIG. \ref{Fig_Dispersion_relation}(h,i), the orbit distributions along the thickness of both counterpropagating modes are very similar; one can expect that the dipolar interaction plays a primary role in the frequency shift across the entire applied magnetic field range. It can be analyzed by focusing on the dynamic dipolar charge differences between counterpropagating SW modes. A difference between such charges also produces a difference in their dynamic dipolar fields. Both fields exert torques on the magnetization that differ from each other. Each torque forces the magnetization to oscillate at a given frequency, therefore a frequency shift is produced once both torques are different from each other. Dipolar dynamic charges can be separated in two types, those with volumetric origin $\rho_v=-\nabla\cdot\mathbf{m}$ and those originated at the system surface $\sigma=\mathbf{m}\cdot\hat{s}$, where $\mathbf{m}=\mathbf{m}_n$ if $y_{n-1}\leq y<y_{n}$ is the dynamic magnetization at the n'th layer with $y_n=nd$, and $\hat s=\hat s_0=-\hat y$ ($\hat s=s_N=\hat y$) is the normal vector at the bottom (top) system surface. In the following, we focus only in the surface charges because, as seen in FIG. \ref{DensimyvsH}(a), the ratio between them ($\sigma[+k_x]/\sigma[-k_x]=m_{Y_0}[+k_x]/m_{Y_0}[-k_x]$) at the bottom surface reflects most of the frequency shift behavior shown in \ref{DensimyvsH}(b). Note that the surface charges at the top surface are not accounted for here since the orbit amplitude at the CoFeB is strongly diminished. As seen, the surface charge ratio behaves similarly to the frequency shift at both spin wave modes. For $0\leq \mu_0 H\leq \mu_0 H_c$ (with $\mu_0H_c\approx 90$) mT, the bilayer is in a homogeneous equilibrium magnetization; consequently, the frequency shift and the surface charge ratio are almost constant. At larger field values, both the surface charge ratio and frequency shift decrease (increase) at the mode 1 (mode 2) as the magnetic field increases, which is consequent to the fact that the equilibrium magnetization state is in a helical configuration where the helical twisting degree of the magnetization increases with the applied field magnitude. A difference between the frequency shift and the surface charge ratio appears at the applied magnetic field where both are zero. It occurs at $\mu_0 H\approx 111$ mT ($\mu_0 H\approx 122$ mT) for the frequency shift (surface charge ratio). This difference can be explained from interlayer exchange interaction. As seen in FIG. \ref{Fig_Dispersion_relation}(h-j), the orbit distribution at $k_x=40$ rads$/\mu$m ($k_x=-40$ rads$/\mu$m) is slightly more (less) homogeneously distributed along the bilayer thickness.  Understanding that more (less) magnetization homogeneity involves less (more) interlayer exchange interaction and that a strengthened interlayer interaction induces a tendency to oscillate at larger frequencies,  one can conclude that the interlayer exchange interaction tends to strengthen the oscillation frequency of spin wave orbit distribution at $k_x=-40$ rads$/\mu$m more than at $k_x=40$ rads$/\mu$m. It tends to move the frequency shift to negative values, leading to $\Delta f=0$ at a lower applied magnetic field than in the case of the surface charge ratio. 

Let's continue with analyzing the frequency shift of the CoFeB(25)/NiFe(47) bilayer. As seen in the upper ($k_x=+40$ rads$/\mu$m) and lower ($k_x=-40$ rads$/\mu$m) panels in FIG. \ref{Fig_Dispersion_relation}(j,k), counterpropagating spin waves of mode 1 have different orbit distributions from each other along the system thickness, one can expect that the interlayer exchange interaction plays the primary role in the frequency shift. In contrast, the orbit distribution of mode 2 along the thickness of counterpropagating spin waves is such that the dipolar and interlayer exchange interactions jointly determine the frequency shift. In the following, we will discuss each mode separately. 

\begin{figure*}[ht!]
\includegraphics[scale=0.19]{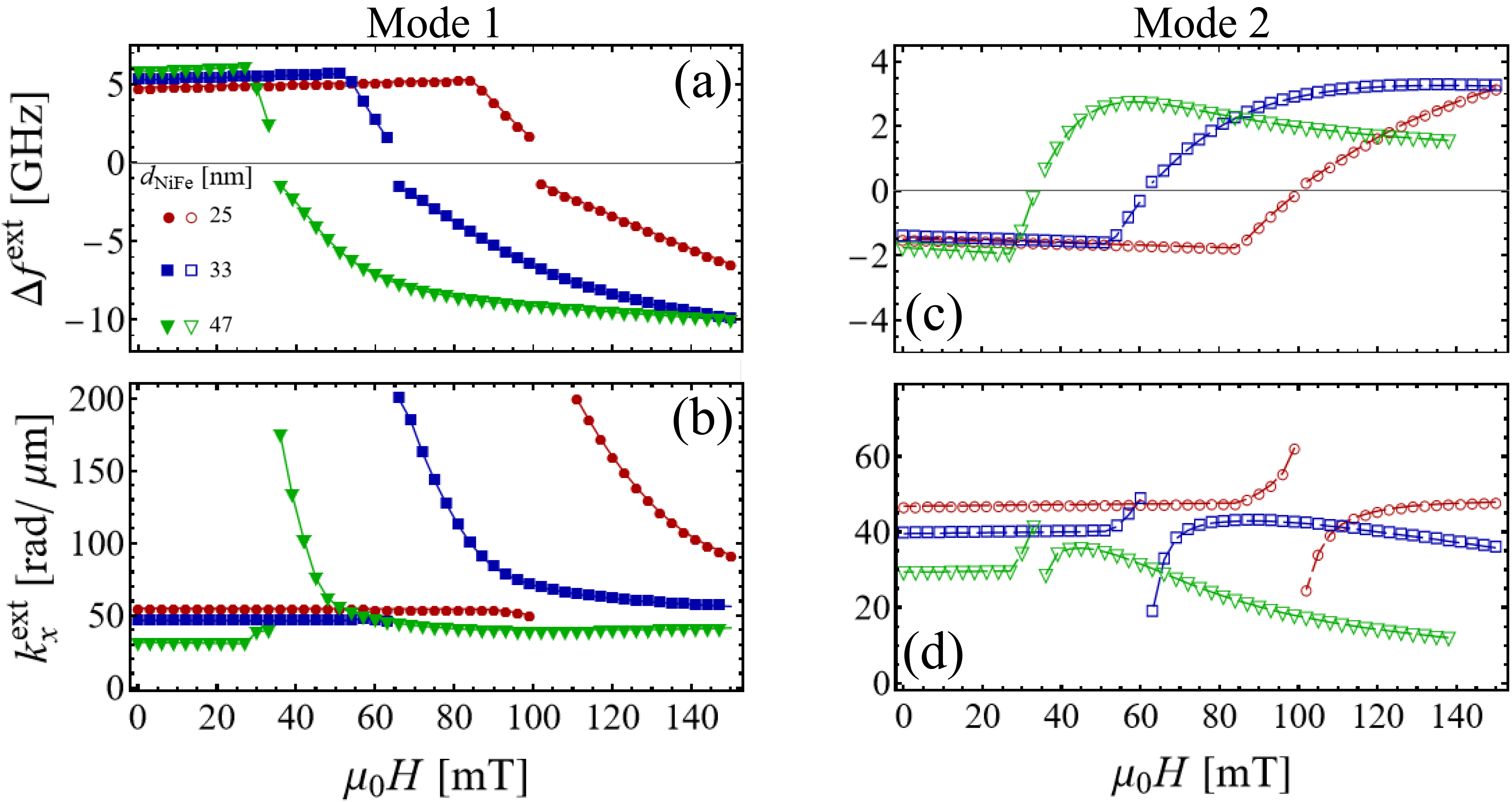}
\caption{(a-c) Frequency shift extreme values $\Delta f^{\text{ext}}$ at (b-d) the corresponding wave vector $k_x^{\text{ext}}$ of a CoFeB/NiFe bilayer, as a function of the applied magnetic field. The CoFeB sublayer thickness is set to $d_{\text{CoFeB}}=25$ nm, whereas three different thicknesses of the NiFe sublayer were considered.}
\label{Helical_States}
\end{figure*}

In the case of mode 1, the influence of the dipolar interaction evaluated via the surface charge ratio (see FIG. \ref{DensimyvsH}(c)) does not explain the frequency shift (see FIG. \ref{DensimyvsH}(d)). This is consequent to the fact that the spin wave orbit distribution along the thickness of the counterpropagating modes has important differences between them, wherein the frequency shift is therefore dominated by the interlayer exchange interaction as aforementioned\cite{negrete2026exchange}. One can argue that modes with greater inhomogeneity in the orbital distribution exhibit a stronger interlayer exchange interaction, which promotes a higher oscillation frequency. In this sense, the frequency shift will be positive (negative) if the orbit distribution along the thickness at $k_x=40$ rads$/\mu$m ($k_x=-40$ rads$/\mu$m) is more inhomogeneous than in the case of $k_x=-40$ rads$/\mu$m ($k_x=40$ rads$/\mu$m). We use this argument to understand the frequency shift of mode 1 as a function of the applied field. Accordingly, one can see that at $H=0$ in FIG. \ref{Fig_Dispersion_relation}(j) - downward red triangle - and $k_x=40$ rads$/\mu m$ (upper figure), the orbit distribution is more inhomogeneous than in the case of $k_x=-40$ rads$/\mu m$ (bottom figure), therefore the frequency shift is positive as shown in FIG. \ref{DensimyvsH}(d). This tendency remains as far as $0\leq H\leq H_C$ with $\mu_0 H_C\approx 27$ mT, wherein the bilayer is in a homogeneous magnetization state. At $H>H_C$, the equilibrium magnetic state turns into a helical configuration, wherein the frequency shift start to displace towards negative values as the twist of the helical equilibrium state increases. It is consistent with the modes distribution at $\mu_0 H\in\{90, 111, 150\}$ mT, wherein the spin wave orbit distribution at $k_x=-40$ rads$÷\mu$m is more inhomogeneous along the thickness than those at $k_x=40$ rads$/\mu$m, letting the spin wave modes at $k_x=-40$ rads$/\mu$m oscillating at larger frequencies than at $k_x=40$ rads$/\mu$m, resulting in a negative frequency shift. 

In mode 2, the frequency shift is explained as a compensation between the dipolar and interlayer exchange interactions. As seen in FIG. \ref{DensimyvsH}(c) of mode 2, the surface charge ratio can not explain by itself the frequency shift shown in  FIG. \ref{DensimyvsH}(d). For a successful explanation, it is necessary to introduce in the analysis the effects of the interlayer exchange interaction via the inhomogeneity of the orbit mode along the thickness. At $H=0$, one can see that the orbit distribution at $k_x=-40$ rads$/\mu$m (see bottom figure at FIG. \ref{Fig_Dispersion_relation}(k) - downward empty blue triangle) is slightly more inhomogeneous than the counterpropagating one at $k_x=40$ rads$/\mu$m (see upper figure at FIG. \ref{Fig_Dispersion_relation}(k) - downward empty blue triangle). It suggests that the spin wave at $k_x=-40$ rads$/\mu$m oscillates at a higher frequency, thereby promoting a negative frequency shift. In contrast, the positive surface charge ratio promotes a positive frequency shift. Since the orbit inhomogeneities at $k_x=-40$ rads$/\mu$m are slightly larger than at $k_x=-40$ rads$/\mu$m, the negative frequency shift promoted by the interlayer interaction is not dominant. As a result, both interactions compensate each other, reducing the frequency shift to $\Delta f\approx -1$ GHz, which is an indication that at $H=0$ the interlayer exchange interaction has a stronger effect on the frequency shift than the dipolar interaction. This negative frequency shift remains almost unaltered as long as the equilibrium magnetization remains in the homogeneous configuration. Once the applied magnetic field exceeds $H_c=27$ mT, the equilibrium state becomes a helicoidal configuration, and the frequency shift of mode 2 becomes positive, as seen in FIG. \ref{DensimyvsH}(d). This is consistent with the counterpropagating spin wave orbit distribution at $\mu_0 H\in\{90, 111, 150\}$ mT. At these fields, the surface charge ratio increases negatively in the range $-1<\sigma[+|k_x|]/\sigma[-|k_x|]<0$, thus promoting a negative frequency shift as the applied field increases. However, the spin wave orbit distribution at $k_x=+40$ rads$/\mu$m acquires a slightly less inhomogeneous profile than the orbit distribution at $k_x=-40$ rads$/\mu$m, promoting a reduced positive frequency shift if compared with $H=0$. We can expect that this positive frequency shift should remain almost unaffected in the set of fields $\mu_0 H\in\{90, 111, 150\}$ mT, since the spin wave orbit distributions slightly change at these fields (see FIG. \ref{Fig_Dispersion_relation}(k)) and with it, a slight change in the interlayer interaction. Afterward, we ended up with a dipolar interaction promoting an increasingly negative frequency shift and a reduced, but almost unaffected, promotion of the positive frequency shift by the interlayer exchange interaction. As a result, both interactions compensate each other, showing a decreasing frequency shift to almost zero at $\mu_0 H= 150$ mT, as shown in FIG. \ref{DensimyvsH}(d) mode 2.

Finally, we characterized the extreme values of the frequency shift $\Delta f^{\text{ext}}$ at its wave vector $k_x^{\text{ext}}$ at three different NiFe thicknesses in the CoFeB(25)/NiFe bilayer, for an applied magnetic field ranging as $0\leq \mu_0 H\leq 150$ mT.  FIG. \ref{Helical_States}(a,b) and (c,d) show $\Delta f^{\text{ext}}$ and $k_x^{\text{ext}}$ as functions of the applied magnetic field of mode 1 and mode 2, respectively. The discontinuity (or gap) in $\Delta f^{\text{ext}}$ of mode 1 at FIG. \ref{Helical_States}(a) are extremes with wave vectors $k_x^{\text{ext}}$ located beyond the wave vector range considered in our calculations. The almost flat behavior of $\Delta f^{\text{ext}}$ and $k_x^{\text{ext}}$ shown in FIG. \ref{Helical_States}, occurs as far as the equilibrium magnetization is in the homogeneous configuration. It happen at applied fields below the coercive field, beyond which the helical magnetic state is stabilized. It is worth noticing the frequency shift values obtained here, reaching the $|\Delta f\approx 10| $ Ghz at large wave vectors $k_x^{\text{ext}}\approx 50$ rads$/\mu$m, or relatively large frequency shifts $|\Delta f|\approx 5$ Ghz at very large wave vectors $k_x^{\text{ext}}\approx 100$ rads$/\mu$m (or sub-100 nm wave lengths). Large wave vectors with large frequency shifts can be tuned by selecting the appropriate NiFe sublayer thickness; however, this is the subject of a forthcoming work. 

Finally, the frequency-shift mechanism proposed throughout this manuscript, based on differences in the dynamic dipolar and interlayer exchange interactions arising from distinct spin-wave mode profiles across the film thickness for two counterpropagating modes at the same wave vector, contrasts with recent literature, which attributes the shift solely to dipolar interactions. We argue that such an interpretation is incomplete.\cite{negrete2026exchange}

\section{Conclusions}

In this work, we have extended the dynamic matrix formalism to describe the spin-wave propagation in a heterostructured multilayer with an inhomogeneous in-plane helical equilibrium magnetization. We have validated our model by fitting experimental and simulated dispersion relations for a SmCo/Fe exchange spring bilayer (see Jiang et al. (2025) \cite{l93m-gb54}), as shown in Supplementary Material Section II \cite{SupMat}. Instead of focusing on SmCo/Fe, our system of interest was the CoFeB/NiFe bilayer due to its versatile features as a magnonic diode and an exchange spring. In contrast to recent literature where the frequency shift behavior is attributed solely to the dipolar interaction, we explained its features by analyzing the differences in the dynamic dipolar and interlayer exchange interactions arising from the distinct spin-wave mode profiles across the thickness for counterpropagating modes at the same wave vector $k_x$. Complimentarily, we have evaluated the extremes frequency shifts $\Delta f^{\text{ext}}$ at the given wave vector $k_x^{\text{ext}}$, finding a combination of large frequency shift values and sub-100 nm spin waves wave lengths that can be tuned with the twisting degree of the helical magnetization state by the application of the external field, and with the thickness of the NiFe sublayer, which might be highly relevant for magnonic applications. 

\section*{Declaration of competing interest}
The authors declare that they have no known competing financial interests or personal relationships that could have appeared to influence the work reported in this paper.

\begin{acknowledgments}
  This research has received funding support from Chilean Doctorado Nacional ANID via fellowship Grant 21211429..
\end{acknowledgments}

\section*{Data Availability Statement}

The data are available upon reasonable request from the authors.

\section*{CRediT authorship contribution statement}
 \textbf{Claudia Negrete}: Writing - original draft, Visualization, Conceptualization, Investigation, Formal Analysis. \textbf{Omar J. Suarez}: Writing - review . \textbf{Attila K\'akay}: Writing - review \&
editing, Conceptualization, Investigation. \textbf{Jorge A. Ot\'alora}: Writing - review \& editing, Conceptualization, Supervision, Resources, Project administration, Methodology, Investigation, Formal Analysis.

\bibliographystyle{apsrev4-2}
\bibliography{References}

\end{document}